\documentclass[twocolumn]{revtex4}

\begin{document}

\title{Theory of phase fluctuating d-wave superconductors and the
spin response in underdoped cuprates}

\author{Igor F. Herbut and Dominic J. Lee}

\address{Department of Physics, Simon Fraser University, 
Burnaby, British Columbia, Canada V5A 1S6 }

\begin{abstract}
The minimal theory of spin of
gapless quasiparticles coupled to fluctuating
vortex defects in the phase of the d-wave superconducting order parameter
at $T=0$ is studied. We find a single superconductor-spin density wave
 phase transition,
which may be fluctuation induced first-order, at which vortices condense
and the chiral symmetry for fermions dynamically breaks.
We compute the spin-spin correlation function in the 
fluctuating superconducting state, and discuss some prominent trends 
in the neutron scattering data on underdoped cuprates in
light of our results.
\end{abstract}
\maketitle

The underdoped high temperature superconductors are highly anisotropic,
quasi two-dimensional materials, in which thermal and quantum
fluctuations of the phase of the superconducting order parameter (OP)
should be important well below the pseudogap temperature
\cite{schneider}. It has recently been shown that a  
phase fluctuating d-wave superconductor (dSC) at zero temperature  ($T=0$)
is inherently unstable towards the formation of the insulating
spin density wave (SDW) state with the loss of phase coherence \cite{herbut},
\cite{babak}. This result
follows from the realization that the effective low-energy theory for the
gapless d-wave quasiparticles, besides the usual spatial
symmetries, also possesses an additional
internal ("chiral") symmetry. This symmetry
becomes dynamically broken when the superconducting phase
coherence is lost via proliferation of vortex defects.
The effective theory of the phase fluctuating dSC in this formulations
is  closely related to
the three dimensional quantum electrodynamics ($QED_3$) \cite{franz},
in which the  "charge" is proportional to the vortex condensate \cite{herbut},
\cite{franz}. In first approximation the fluctuations of the gauge-field
may be completely neglected in the superconducting state, which 
yields quasiparticles as well defined excitations. When the vortices
condense and the charge in the effective $QED_3$ becomes finite, the
chiral symmetry of the dSC becomes dynamically broken \cite{pisarski}.
In the present context this translates into the SDW ordering,
with confined spin-1/2 (spinon) excitations \cite{herbut}.

  Several issues of direct relevance for underdoped cuprates
  naturally arise in such a theory of the
 phase fluctuating dSC. Can the SC and the
 SDW long-range orders coexist? Several intriguing
 recent experiments \cite{sonier}, \cite{mook}, \cite{sidis}, find
 possible signs of such 
 coexistence of the two inimical orderings. If there is a single
 dSC-SDW quantum phase transition, on the other hand,
 what should be its universality class? What are the effects of
 the vortex fluctuations {\it inside} the superconducting state?
 These questions all call for a better understanding of the
 quasiparticle-vortex interaction, particularly in the superconducting state.
 In this Letter we investigate probably the simplest 
 theory of the low-energy d-wave quasiparticles
 coupled to the complex scalar vortex field (dual to the 
 superconducting OP). The theory describes the decoupled
 spin sector of the phase coherent dSC at $T=0$
and subject to strong quantum vortex fluctuations.
 In underdoped high temperature superconductors such fluctuations
 should presumably arise from the
 Coulomb interaction that becomes effectively
 stronger towards half filling, and tends to disorder the phase of
 the superconducting OP. If extended to the non-superconducting phase,
 our theory
 reduces to the previously studied $QED_3$. First, we argue that the 
 condensation of vortices (i. e. the loss of superconductivity) and
 the  breaking of the chiral symmetry for fermions (the
 SDW instability) in our theory coincide. Second, we show  that
 the quantum dSC-SDW phase transition can be either in the modified XY
 universality class, or it could be fluctuation-induced first-order,
 depending on the values of the couplings in the theory. Finally
 and most importantly,  we 
 calculate the spin dynamics {\it induced} by vortex fluctuations deep
 inside the dSC in the leading approximation.  The intricate evolution of the
 spin response with frequency observed in inelastic
 neutron scattering experiments on YBCO \cite{dai},\cite{fong}, \cite{arai} 
 follows quite naturally from our results.
 In particular, we predict the appearance of four weak and narrow "diagonally"
 incommensurate peaks at low energies, the energy of which vanishes  
 with the superconducting $T_c$.

Consider the quantum mechanical ($T=0$) action for the low energy
quasiparticles in the two-dimensional phase fluctuating dSC,
$S= \int d^3 x {\cal L}$, $x = (\tau, \vec{r})$, and
\begin{eqnarray}
{\cal L} = \sum_{i=1}^N
\bar{\Psi}_i \gamma_\mu (\partial_\mu - i a_\mu) \Psi_i
+ \frac{i}{\pi} \vec{a} \cdot (\nabla \times \vec{A}) + \\ \nonumber
| (\nabla - i \vec{A} )\Phi|^2 + \mu ^2 |\Phi|^2 + \frac{b}{2} |\Phi|^4.
\end{eqnarray}
Here the fluctuating complex field $\Phi$ describes the vortex loops,
and $\langle\Phi \rangle = 0$ implies the superconducting phase coherence.
Two ($N=2$) four-component Dirac fermions describe
the gapless, neutral, spin-1/2 (spinon) excitations near the four nodes
of the superconducting order parameter, one Dirac field for each pair of
diagonally opposed nodes \cite{herbut}.  $\vec{a}$ is the  gauge-field 
that results from properly
absorbing the singular part of the superconducting phase due to vortices
into the spinon fields \cite{franz}, \cite{herbut}.
$\gamma_\mu$, $\mu =0,1,2$ are the Dirac gamma matrices
\cite{herbut}. Finally, $\vec{A}$ is an additional
auxiliary (Chern-Simons) field which facilitates the statistical
spinon-vortex coupling, as will be explained shortly. The single 
tuning parameter $\mu^2$ may be understood as related to doping, $\mu^2 \propto (x-x_c)$, with
$x_c$ being the critical doping in the underdoped regime.
The coupling $b$ describes the short-range repulsion between the vortex loops.
In our units $\hbar=c=v_F=v_\Delta=k_B =e=1$.

We may motivate the field theory (1) as follows \cite{dlee}:
the integration over the gauge-field $\vec{a}$ leaves the spinon-vortex coupling
$i \vec{A} \cdot \vec{J}_{\Phi}$, where $\nabla\times \vec{A}= \pi 
\vec{J}_{\Psi}$, and $\vec{J}_{\Psi}$  and
$\vec{J}_{\Phi}$ are the spinon and vortex current densities,
respectively. Vortices and spinons therefore see each other as
sources of magnetic flux, and circling around a vortex with a spinon
(or vice versa) leads to a phase change of $\pi$. This is precisely
the {\it statistical} (Aharonov-Bohm) part of the interaction between vortices and
spinons \cite{franz}, \cite{herbut}, \cite{dlee}. Without fermions ($N=0$), the
integration over the gauge fields leads to the standard $|\Phi|^4$ theory,
in the universality class of the XY model. Strictly speaking, for $N=0$ the
theory should reduce to the Higgs scalar electrodynamics, dual to
the $|\Phi|^4 $ theory. This can be accomplished by introducing an
additional fluctuating gauge field into the theory that
would mediate a long-range interaction between vortices and represent
the electrical charge. Following the previous work \cite{herbut},
\cite{franz}, \cite{dhlee}
we will assume spin and charge to be separated at low
energies, so we may neglect the charge degrees of freedom. This
assumption is surely justified inside the superconducting state 
which is the subject of the present work. For subtle issues
that arise in the normal state the reader is referred to \cite{igornew}.
Finally, the velocity anisotropy ($v_F \gg v_\Delta$)
and all the local interactions between quasiparticles \cite{herbut}
have been neglected, as irrelevant at low
energies \cite{lee}, \cite{vafek}.

  Let us discuss first how the theory (1) is related to the $QED_3$
\cite{herbut}, \cite{franz},
in the simplest mean-field approximation for the vortex
field. Neglecting the fluctuations in $\Phi$, in the superconducting
phase ($\langle \Phi \rangle = 0 $), the integration over $\vec{A}$
constrains $\nabla \times \vec{a} =0$, and spinons are essentially free.
When vortices condense ($\langle \Phi \rangle \neq 0$), on the other
hand, the gauge-field $\vec{A}$ acquires a mass via Higgs mechanism,
$|\langle \Phi \rangle| ^2 \vec{A}^2 $. The Gaussian integration over
$\vec{A}$ produces then the Maxwell term for $\vec{a}$,
$\sim (\nabla \times \vec{a})^2 / |\langle \Phi\rangle| ^2$. Together 
with the Dirac Lagrangian this constitutes the 
$QED_3$ for spinons, in which the fermions' chiral symmetry generated
by $\gamma_3$ and
$\gamma_5$ is dynamically broken by the generation of the mass term 
 $\sim M \bar{\Psi}_i \Psi_i$ \cite{pisarski}. In the present context such a
mass $M\sim \langle {\bar \Psi}_i \Psi_i\rangle \sim |\langle \Phi \rangle|^2 $
is proportional to the SDW OP, with the (incommensurate) ordering
wave vectors that connect the diagonally opposed nodes
of the superconducting OP \cite{herbut}.  Carefully 
including the charge degrees of freedom in (1) 
one also finds the SDW to be an electrical insulator \cite{igornew}. 

In the mean-field approximation, the vortex condensation and the
SDW transition in the theory (1) coincide.
It is not obvious that this feature survives the
inclusion of the fluctuations, and it is conceivable
that the SDW transition  may occur within the
superconducting phase \cite{herbut}, \cite{pereg}.
To examine this issue we
first ask what would be the {\it exact} propagator for the gauge field
$\vec{a}$ if there were no fermions, i. e. when $N=0$ in (1).
We then approximate the interacting  
action for $\vec{a}$ that would result from the integrations over $\Phi$ and
$\vec{A}$ with the effective Gaussian term that reproduces that exact propagator.
In the Landau gauge such a propagator is 
\begin{equation}
G^0 _{aa,\mu \nu} (p,m) = \frac{\pi^2 \Pi_{AA} (p,m)}{p^2}
(\delta_{\mu \nu}-\hat{p}_\mu \hat{p}_\nu ),
\end{equation}
where $\Pi_{AA} (p,m) (\delta_{\mu \nu}-\hat{p}_\mu \hat{p}_\nu )$
is the transverse current-current correlation function in the
$|\Phi|^4$ theory. Here $m^2 =\mu^2 +O(b)$ is
the fully renormalized "mass" of
the vortex field. In general, $\Pi_{AA}(p,m) = n p F_ + (m/p)$,
for $m^2 >0$ (superconductor). To the lowest order in $b$
\begin{equation}
F_+ (z) = \frac{1}{8\pi} [ (4z ^2 +1) \arctan(\frac{1}{2z}) - 2z] +O(b).
\end{equation}
Here  we generalized our model to the one with $n$ complex
vortex fields, $n=1$ being the case of physical interest.

 We are now in position to
  study the chiral symmetry breaking for fermions with (2)
 serving as the bare (without fermion polarization) gauge-field
 propagator in the
 $QED_3$. Consider the standard large-N Dyson equation \cite{pisarski}
 for the fermion self-energy
 \begin{equation}
 \Sigma(q) = \frac{1}{4} Tr \int \frac{d^3 p}{(2\pi)^3}\gamma_\mu
 \frac{ G_{aa, \mu \nu} (p-q ,m) \Sigma(p)}{ p^2 + \Sigma(p)^2 }
 \gamma_\nu
 \end{equation}
 Here $[G _{aa,\mu \nu} (p,m)]^{-1} = \Pi^F _{aa,\mu \nu}(p) +
 [G^{0} _{aa,\mu \nu} (p,m)]^{-1}$,
 with the one-loop fermion polarization $ \Pi^F _{aa,\mu \nu}(p)= 
 (Np/8)(\delta_{\mu \nu} - \hat{p}_\mu \hat{p}_\nu)$.
 It is useful to consider first the point of the 
 superconducting phase transition $m=0$, where Eq. (3) implies 
 $\Pi_{AA} (p,0) =  np(1 /16 +O(b_c ) )$, where $b_c$ is the
 fixed point value of $b$ in $d=3$ \cite{cha}. Linear dependence
 of $\Pi_{AA}(p,0)$ on momentum is an exact result
 \cite{herb-tes}. Inserting this into the Dyson equation  we find
 that at $m=0$ the effect of vortices is only to increase  
 the coefficient $N\rightarrow N_{eff}=N+ 128/(\pi^2 n) + O(b_c)$.
 Recalling that the non-trivial solution of the Dyson equation (4)
 exists only for $ N_{eff} < N_c = 32 / \pi^2 $ \cite{pisarski},
 for $N=2$ we find that chiral symmetry
 at $m=0$ would be already broken only for
  $n > n_c = 10.44( 1+ 32/(9\pi^2 n ))$,
  where we have also included the known
  $O(b_c)$ correction to $\Pi_{AA}(p,0)$
 in the large-$n$ approximation \cite{cha}.
 Since $n=1$ in the physical case we conclude that right at the
 superconducting critical point SDW order is most likely absent.

 Little further thought shows that the above result implies that the 
 SDW order
 is absent for all $m^2 >0$. Indeed, this is to be expected, since
 $G^0 _{aa} (p,m)< G^0 _{aa} (p,0)$, and
 $\vec{a}$ is only stiffer in the superconducting
 phase. To prove this, assume that for $m^2 >0$ a
 non-trivial solution of the Eq. (4), $\tilde{\Sigma}(q)$, does exist.
 Such a solution would then satisfy
 \begin{equation}
 \tilde{\Sigma}(q) <  \frac{1}{4} Tr \int \frac{d^3 p}{(2\pi)^2}\gamma_\mu
 \frac{ G_{aa,\mu \nu} (p-q,0) \tilde{\Sigma}(p)}
 { p^2 + \tilde{\Sigma} (p)^2} \gamma_\nu.
 \end{equation}
 On the other hand, we already established that there is only a trivial
 solution of the Dyson equation at $m=0$. This means that assuming a
 candidate function $\Sigma(p)$ and inserting it under the integral
 in the Dyson
 equation (4) for $m=0$ will produce only a {\it smaller} new $\Sigma(q)$,
 since under iterations the physically acceptable self-energy must approach
 the trivial solution. This  further implies that any
 $\Sigma(q)$ from the domain of attraction of the trivial solution
 at $m=0$ has to satisfy the inequality opposite to (5). 
 A non-trivial $\tilde{\Sigma}(q)$ at $m^2 >0$ therefore can not
 exist if it did not exist at $m^2 =0$.

   For $m^2 <0$ superconductivity is lost, and $\Pi_{AA} (p,m)
 = n p F_- ( |\langle \Phi \rangle|^2 /p )$, with $F_-(z)
 \rightarrow 1/16$ for $z \ll 1$, and $F_- (z)=2 z$, for $z \gg 1$
(Higgs mechanism), to the leading order.
Since $G^0 _{ aa} (p,m)> G^0 _{ aa} (p,0)$
 for $m^2 <0$, the above argument no longer applies. In fact, it is easy to
 see that there is {\it immediately}
  a non-trivial solution of the Eq. (4) when
 $m^2 < 0$. Since $\Sigma(q)>0$ only for  $ q \sim |\langle \Phi \rangle|^2$
 and smaller, $|\langle \Phi \rangle|^2 $ serves as the effective ultraviolet
 cutoff in the Eq. (4). The Dyson equation then reduces to the
 standard one in the $QED_3$
 \cite{pisarski}. For $N=2 < N_c $,  $\Sigma(q=0) \sim
 |\langle \Phi \rangle|^2$, and the chiral symmetry is dynamically broken.
 There is no intermediate (quantum disordered)
 phase in between the SDW and the dSC in the theory (1), unless the exact
 value of $N_c $ in the $QED_3$ is actually less than two \cite{cohen}.

   The above argument strongly
   suggests that there is a single dSC-SDW transition
 in the theory (1), but does not say anything about its nature. To
 study this issue it is better to proceed in the opposite direction,
 and integrate over the fermions first. This gives dynamics to $\vec{a}$
 via the fermion polarization bubble.
 If we neglect the quartic and the higher
 order terms and perform the Gaussian integral over $\vec{a}$, the result is
 the Maxwell-like term for $\vec{A}$:
 $\langle A_\mu (p)A_\nu(-p) \rangle = (N\pi^2 /8|p|)
 (\delta_{\mu\nu}- \hat{p}_\mu \hat{p}_\nu) $.
 Note that since the inverse of the
 above average is non-analytic at $p=0$,
 it can not renormalize \cite{herbut1}, and therefore the number of fermions
 $N$ represents an exactly {\it marginal} coupling. 
 Assuming a constant $\Phi$ in (1) we may further integrate $\vec{A}$
 to find the energy per unit volume to be
 \begin{eqnarray}
 S[\Phi] - S[0] = (\frac{\mu^2}{\Lambda^2} + \frac{N}{48})\Phi^2 +
 \frac{1}{2}(\frac{b}{\Lambda}- \frac{\pi^2 N^2}{48}) \Phi^4 + \\ \nonumber 
 + \frac{1}{6\pi^2} \ln( 1+ \frac{\pi^2 N}{4}\Phi^2)  + 
 \frac{\pi^4 N^3}{384}\Phi^6 \ln (1+ \frac{4}{\pi^2 N \Phi^2}), 
 \end{eqnarray}
 where we have rescaled $\Phi^2 /\Lambda \rightarrow \Phi^2$,
 with $\Lambda$ being the ultraviolet cuttof.
 Standard analysis of the Eq. (6) shows that there is a discontinuous
 transition for $ N > (4/\pi)\sqrt{2b/\Lambda}) $. Using the lowest order
 fixed point value $b/\Lambda = (2\pi^2 /5)(4-d) + O( (4-d)^2 )$
 in $d=3$ as a crude estimate of this bound,  we find the first-order
 transition for $N > 3.58 $, in rough agreement with
 \cite{dlee}. Of course, this conclusion is to be trusted only
 for $N\gg 1$, when the first-order transition occurs at large
 $\mu^2$, at which our neglect of fluctuations in $\Phi$ in arriving
 at the Eq. (6) becomes
 justified. For $N\ll 1$, on the other hand, the phase transition
 in the theory (1) should be continuous, with weakly modified XY exponents:
 $\nu=\nu_{xy}+O(N)$, $\eta=\eta_{xy}+O(N)$ \cite{dlee}. The situation is
 reminiscent of the Ginzburg-Landau superconductor 
 \cite{herb-tes}, with $N \gg 1$ analogous to the strongly type-I,
 and $N\ll 1$ to the extreme type-II case.

 Finally, we turn to spin dynamics deep inside the dSC, when $m/b \gg 1 $.
 For small momenta, $p \ll m$, we find then that 
 $G^0 _{aa,\mu \nu} (p,m) = ( (\pi/24 m) +O(p^2) )
 (\delta_{\mu\nu}- \hat{p}_\mu \hat{p}_\nu ) $. Integrating out such a 
 {\it massive} $\vec{a}$  leads to the effective $2+1$ dimensional 
 Thirring model  for fermions in the dSC 
 \begin{equation}
{\cal L} = \bar{\Psi}_i \gamma_\mu \partial_\mu  \Psi_i + 
\frac{\pi}{72 m} \bar{\Psi}_i \gamma_\mu \Psi_i
\bar{\Psi}_j \gamma_\mu \Psi_j,
\end{equation}
with the summation over repeated indices assumed. Using the
Hubbard-Stratonovich transformation we can rewrite this as
\begin{eqnarray}
{\cal L} = \bar{\Psi}_i \gamma_\mu \partial_\mu  \Psi_i
+ \frac{18 m }{\pi} Tr [ M_{\mu} ^{ij} M_{\mu}^{ji}]  + 
Tr [ M_{\mu} ^{ij}\gamma_\mu \Psi_j \bar{\Psi}_i ]  
\end{eqnarray}
Making an {\it ansatz} $ M_{\mu}^{ij}(x) =
(M(x)/3) \delta_{ij} \gamma_\mu$ at the saddle point,
and then integrating out the fermions yields
\begin{equation}
S= N\int d^3 x [ \frac{24 m}{\pi} M^2 (x)
- Tr [ ln ( \gamma_\mu \partial_\mu - M(x) ) ]   ]. 
\end{equation}
Expanding further in powers of $M(x)$ we finally write
\begin{equation}
\frac{S}{N}= \int \frac{d^3 q}{(2\pi)^3 }
( \frac{24 \pi m- \Lambda}{\pi^2} + \frac{|q|}{8} )M^2 (q) + O(M ^4 ), 
\end{equation}
where $M^2 (q) = M(q) M(-q)$.
On the other hand, using the definition of the Dirac fields in terms
of the electron creation and annihilation
operators \cite{herbut}
\begin{equation}
\langle M(x) \rangle =
\frac{\pi}{12 Nm} \langle \hat{S}_z ( \vec{r},\tau) \rangle
\sum_{i=1}^2 \cos( 2 \vec{K}_i \cdot \vec{r}), 
\end{equation}
 where the vectors $\pm \vec{K}_i$, $i=1,2$ denote the positions of the four
nodes of the d-wave order parameter. $\langle M(q=0)\rangle $ is
therefore the static SDW OP \cite{herbut}. Similarly, 
\begin{eqnarray}
\langle M^2 (\vec{q}, \omega) \rangle
- \frac{\pi}{48 Nm}=  \\ \nonumber 
( \frac{ \pi }{24N  m})^2 \sum_{i=1}^2
( \langle \hat{S}_z ( \vec{q} + 2\vec{K}_i, \omega)
\hat{S}_z ( -\vec{q} -2 \vec{K}_i, -\omega)\rangle +  \\ \nonumber
\langle \hat{S}_z ( \vec{q} - 2\vec{K}_i, \omega)
\hat{S}_z ( -\vec{q} + 2 \vec{K}_i, -\omega)\rangle).  
\end{eqnarray}
Since the dSC is rotationally invariant, the spin-spin correlation function
 is diagonal, $\langle \hat{S}_\alpha ( \vec{k}, \omega)
\hat{S}_\beta ( -\vec{k}, -\omega) \rangle = \chi (\vec{k},\omega)
\delta_{\alpha \beta}$. We may therefore finally deduce that the imaginary
part of the spin response function $\chi(\vec{k},\omega)= \chi ' (\vec{k},\omega)
+ i \chi ''(\vec{k},\omega)$
in the phase fluctuating dSC is
\begin{equation}
\chi '' (2\vec{K}_i\pm \vec{q},\omega ) = (\frac{ 12 Nm}{\pi})^2
Im \langle M^2 (\vec{q}, \omega)\rangle. 
\end{equation}
Analytically continuing to {\it real} frequencies $i\omega\rightarrow
\omega $, in the Gaussian approximation to Eq. (10) we finally obtain
\begin{equation}
\chi '' (2\vec{K}_i\pm \vec{q},\omega )  =
(\frac{24 m}{\pi})^2  \frac{ N \Theta(\omega^2 - q^2)
\sqrt{\omega^2 - q^2 } }
{ (192 m/\pi)^2 + \omega^2 - q^2}, 
\end{equation}
where we have taken $m \gg \Lambda$ deep inside the dSC. With 
$m\rightarrow \infty$ Eq. (14) reduces correctly
to the non-interacting limit, in agreement with \cite{rantner}. 

First, let us connect the uniform vortex susceptibility $m$ to 
the measurable superconducting $T_c$.
Assuming a continuous dSC-SDW transition,
$m^2 \propto (\mu^2 -\mu_c ^2)^{\gamma}$ ,
with $\gamma= \nu (2-\eta)$ characterizing the quantum critical point;
also, $T_c \propto (\mu^2 - \mu_c ^2 ) ^{z \nu}$. Since the dynamical
critical exponent in the theory (1) is $z=1$, 
$m\propto T_c ^{1- (\eta/2)}$. Sufficiently away from the critical
point we may neglect the anomalous dimension $\eta$ 
and find the result dictated by the (engineering)
dimensional analysis, $192 m/\pi = c T_c$, where $c$ is a
number.

 Next, we look for the position of the maximum of the spin response
 function $\chi''(\vec{q},\omega)$
 at a fixed energy $\omega$. Eq. (14) implies that for
 $\omega \ll c T_c$, the maximum values are
  located at four "diagonally" {\it incommensurate}
 wave vectors $\pm 2\vec{K}_{1,2}$ (i. e. at $\vec{q}=0$). As frequency
 increases the peak intensity grows, and recalling that $v_F \gg v_\Delta$
 and taking the lattice periodicity into account 
 one finds the four peaks overlapping first at four "parallel"
 incommensurate positions \cite{rantner}.
 With the further increase of frequency, at some point four initial peaks
 start overlapping at $(\pi,\pi)$, and for a while the "commensurate" response
 dominates.  At $\omega >c T_c$ the maximum in Eq. (14) shifts to a
 $|\vec{q}| \neq 0$, which implies a weak redistribution of the
 commensurate peak to four "parallel" incommensurate positions at largest
 frequencies.
 The predicted evolution of the spin response is in qualitative 
 agreement with the observations in YBCO \cite{arai}. More detailed
 and quantitative picture will be presented in a future publication. 

 Consider then  the commensurate response at $\vec{k}=(\pi,\pi)$,
 that in our notation corresponds to some $\vec{q}\neq 0$.
 As a function of frequency, at $T=0$ $\chi ''$ vanishes 
 below some cutoff energy $\omega_c = |\vec{q}|$. Also, the maximum response
 is at the energy $\omega_0 = \sqrt{ |\vec{q}| ^2 +(c T_c)^2} $,
 which {\it decreases} with decreasing $T_c$, as observed
 \cite{dai}, \cite{fong}. As $T_c\rightarrow 0$,
 however, the commensurate
 peak energy $\omega_0 \rightarrow |\vec{q}| \neq 0$.
 Such a finite $\omega_0$ ($\approx 20meV $) as $T_c \rightarrow 0$ is
 in agreement with the data 
 in YBCO (Fig. 29 in the second reference \cite{dai}). 
 Remarkably, the cutoff energy 
 that can be estimated from a different
 set of data on YBCO (Fig. 17 in \cite{fong} )
 is also $\omega_c\approx 20 meV$, consistent with our prediction.
 At large energies, $\chi ''$ should behave as $\sim 1/\omega$,
 which also appears to be in general accord with the data.

 Finally, at the incommensurate ($\vec{q}=0$) wavevector, $\chi ''
 (2\vec{K}_i,\omega)\sim \omega $ at small energies, with the peak
 at $\omega_0 = c T_c $. Whereas the energy of the resonance
 should go to a finite value as $T_c$ vanishes with underdoping, we
 predict that the energy of the "diagonally" incommensurate peaks, which
 should become discernible at lower energies, should
 extrapolate to zero. Detection of such a soft mode
 inside the dSC state would provide the smoking gun
 evidence for (1) as the low-energy theory for spin of underdoped cuprates.

 This work was supported by NSERC of Canada and the Research Corporation.
IFH also thanks Aspen Center for Physics for its hospitality.

\end{document}